%% file: Cantiello.tex
\documentclass[mypaper,7pt,twoside]{CoAst}
\usepackage{epsf,graphicx,fancyhdr,txfonts}
\input{CoAst_liege-proceedingslayo}

\def\rfr{\smallskip\par\noindent
        \hangindent=7truemm
        \hangafter=1}

 \def\mso{\,\mathrm{M}_\odot}
 
 \def\lso{\,{\rm L}_\odot}
 
 \def\kms{\, {\rm km}\, {\rm s}^{-1}}

 \def\simle{\mathrel{\hbox{\rlap{\hbox{\lower4pt\hbox{$\sim$}}}\hbox{$<$}}}}
 \def\simgr{\mathrel{\hbox{\rlap{\hbox{\lower4pt\hbox{$\sim$}}}\hbox{$>$}}}}

 \def\hpc{\mathrm{H}_{\mathrm{P,c}}}
 
 \def\c2{^{12}{\rm C}}
 \def\c3{^{13}{\rm C}}
 \def\n14{^{14}{\rm N}}
 \def\c1213{^{12}{\rm C}/^{13}{\rm C}}
 \def\he3he4{^3{\rm He}/^4{\rm He}}
 
 \def\logt{\log \mathrm{T}}
 
 \def\microt{\xi}
 \def\vsini{\varv_{\mathrm{rot}}\sin i}
 
 \def\vca{\langle{\varv_{c}}\rangle}

\begin{document}
\sf

\chapterCoAst{On the origin of microturbulence in hot stars}
{M.\,Cantiello,N.\,Langer,I.\,Brott,A.\,de Koter\,S. N. Shore, J. S.\,Vink, A.\,Voegler \& S.-C.\,Yoon} 
\Authors{M.\,Cantiello$^{1}$,N.\,Langer$^{1}$,I.\,Brott$^{1}$,A.\,de Koter$^{1,2}$\,S. N. Shore$^{3}$, J. S.\,Vink$^{4}$, A.\,Voegler$^{1}$ \& S.-C.\,Yoon$^{5}$} 

\Address{
 $^1$ Astronomical Institute, Utrecht University \\ Princetonplein 5, 3584 CC, Utrecht, The Netherlands \\
  $^2$ Astronomical Institute Anton Pannekoek, University of Amsterdam, Kruislaan 403, 1098 SJ, Amsterdam, The Netherlands \\
 $^3$ Dipartmento di Fisica ``Enrico Fermi'', Universit\`a di Pisa, via Buonarroti 2, Pisa 56127 and INFN - Sezione di Pisa, Italy \\
  $^4$ Armagh Observatory, College Hill, Armagh, BT61 9DG, Northern Ireland (UK) \\
   $^5$ Department of Astronomy \& Astrophysics, University of California, Santa Cruz, High Street, Santa Cruz, CA 95064, USA \\
              }
\noindent
\begin{abstract}
We present results from the first extensive study of convection zones in the envelopes of hot massive stars, which are caused by opacity peaks
associated with iron and helium ionization. These convective regions can be located very close to the stellar surface.
The region in the Hertzsprung-Russel diagram in which we predict the convection zones and the strength of this convection is in good agreement 
with the occurrence and strength of microturbulence in massive stars. We argue further that convection close to the surface may trigger clumping
at the base of the stellar wind of hot massive stars.

\end{abstract}

\Session{ \one }
\section*{Introduction}
With the introduction
of the so called iron peak in stellar opacities {\em (Iglesias et al. 1992)} a convection
zone appears in the envelope of sufficiently luminous massive main sequence models
{\em (Stothers \& Chin 1993)}. This convective region contains little mass and is usually not discussed 
in the context of stellar evolution calculations.
Here, we mainly focus on the question whether the occurrence of sub-surface convection might be 
correlated with observable small scale velocity fields at the stellar surface and in the stellar wind. 
A similar idea has been used to explain microturbulence in low mass stars {\em (Edmunds 1978)}, in which  
envelope convection zones are extended and reach the stellar photosphere. 
While {\em Edmunds (1978)}  concludes that the 
same mechanism can {\em not} explain microturbulent velocities in O and B stars, 
the iron-peak induced sub-photospheric convection zones in these stars were not yet
discovered at that time.
In fact, we demonstrate below that sub-surface convection may not only
cause photoshperic velocity fields which are observable, but possibly even induce clumping at the base of the stellar wind.
\section*{Method}
We performed a systematic study of sub-surface convection by calculating models of hot stars with a hydrodynamic stellar evolution code (see for example {\em Yoon et al. 2006)}.
The Ledoux criterion is used to determine which regions of the star are unstable to convection,
and the convective velocity is calculated according to the Mixing Length Theory {\em (B\"ohm-Vitense 1958)}.
The opacities in our code are extracted from the OPAL tables {\em (Iglesias \& Rogers 1996)}.
We calculated a grid of non-rotating stellar evolutionary models for initial masses
between $5\mso$ and $100\mso$, at metallicities of Z=0.02, Z=0.008 and Z=0.004, roughly corresponding to 
the Galaxy, the LMC and the SMC, respectively. 
\section*{Sub-surface convective regions}
The iron convective region in the envelope of hot stars corresponds to a peak in the  
opacity at $\logt \simeq 5.3$.  The appearance and properties of this subsurface convective zone 
have been studied. In particular we used our grid of models to map on the HR diagram the average convective velocity
in the upper part of the iron convection zone $\vca$. Our main findings are: 
\begin{itemize}
\item For given luminosity and metallicity, $\vca$ increases with decreasing surface temperature.
The convection zones are located deeper inside the star (in radius, not in mass), 
and the resulting larger pressure scale height leads to larger velocities. 
At solar metallicity and $10^5\lso$ (i.e. roughly at $20\mso$) the velocities
increase from less then 1 $\kms$ at the ZAMS to more than $5\kms$ in the supergiant regime,
where $\vca = 2.5\kms$ is achieved at $T_{\rm eff}\simeq 30\, 000\,$K.
At SMC metallicity, the iron convection zone
is absent at the ZAMS for $ L < 10^5\lso$, and a level of $\vca = 2.5\kms$ is only reached
at $T_{\rm eff}\simeq 20\, 000\,$K.
\item For given effective temperature and metallicity, the iron zone convective velocity
increases with increasing luminosity, as a larger flux demanded to be convectively transported 
requires faster convective motions. 
We found threshold luminosities below which iron convection zones do not occur, i.e.,
below about $10^{3.2}\lso$, $10^{3.9}\lso$, and $10^{4.2}\lso$ for the Galaxy, LMC and SMC, respectively.
\item The iron convection zones become weaker for smaller metallicities,
since due to the lower opacity, more of the flux can be transported by radiation. 
The threshold luminosity for the occurrence of the iron convection zone 
quoted above for $Z=0.02$ is ten times lower than that
for $Z=0.004$. And above the threshold, for a given point in the HR diagram, the convective velocities
are always larger for larger metallicity. 
\end{itemize}
\figureCoAst{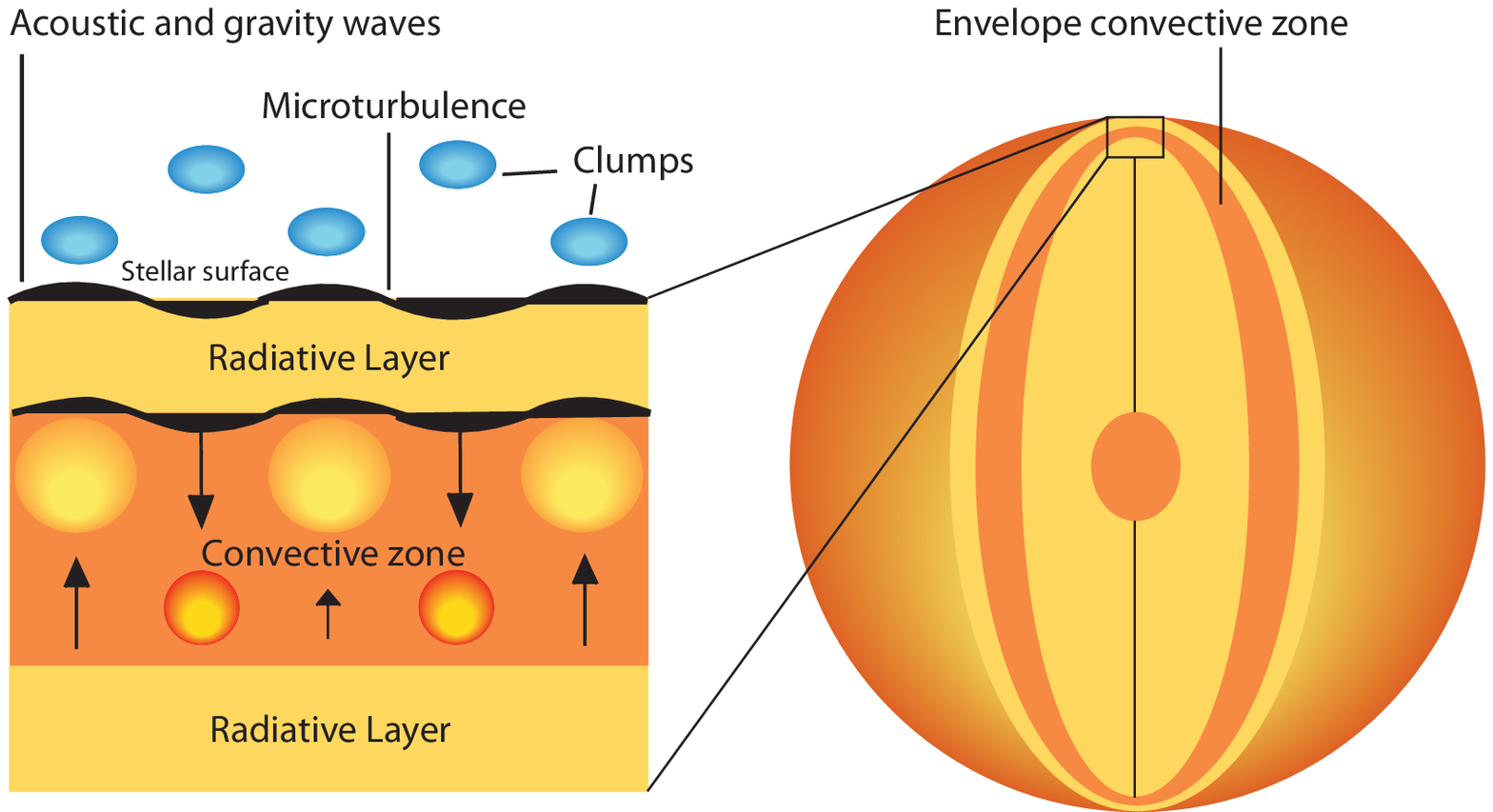}{Schematic representation of the physical processes 
connected to subsurface convection. Acoustic and gravity waves emitted in the convective zone
   travel through the radiative layer and reach the surface, inducing density and velocity fluctuations. 
In this picture microturbulence and clumping at the base of the wind are a
   consequence of the presence of subsurface convection.}{sketch}{t}{clip,angle=0,width=115mm}
\section*{Microturbulence}
The microturbulent velocity  $\microt$ is the non-thermal component of the photospheric velocity field which has a correlation length smaller than
the size of the line forming region. The non-thermal velocity component with a larger correlation length is referred to as macroturbulence  {\em (e.g. Aerts et al., this volume)}.  
In fact the convective cells in the upper part of a convection zone act as pistons and generate 
acoustic and gravity waves propagating outward. The problem of sound waves generated by turbulent motions has been 
discussed by {\em Lighthill (1952)} and extended to a stratified atmosphere by  {\em Stein (1967)} and {\em Goldreich \& Kumar (1990)}. In the presence of stratification, gravity acts as a restoring force and allows the excitation of gravity waves as well. For both, acoustic and gravity waves, the most important parameter determining the emitted kinetic energy flux is the velocity of the convective motions. This is the reason why we used $\vca$  as the crucial parameter determining the efficiency of subsurface convection.
{\em Lighthill (1952)} and {\em Stein (1967)} showed that convection excites acoustic and gravity waves, with a maximum emission for waves with $\lambda \sim \hpc$, the pressure scale height at the top of the convective region. They calculated the amount of convective flux $F_{c}$ that is going into acoustic waves, $F_{\mathrm{ac}}\sim F_{c} M_c^{5}$ , and gravity waves, $F_{{\mathrm g}}\sim F_{c} M_c$, where we take $F_{c}\sim \rho_c \vca^3$ and $M_c$ is the Mach number for convection calculated at the top of the convective region.
Since convection in our models is highly subsonic, gravity waves are expected to extract more energy than acoustic waves from the convective region.  
These waves can propagate outward, steepen and become dissipative in the region of line formation.
Here, they may induce density and velocity fluctuations (Fig.\ref{sketch}). The energy associated with the induced velocity fluctuations must be comparable or smaller than the energy in the waves
above the convective zone. For microturbulence to be excited by this process, it is required that $E_{{\mathrm g}}\geq E_{\microt}$ or
\begin{equation}
\frac{E_{{\mathrm g}}}{E_{\microt}}\sim M_c\left( \frac{\rho_c}{\rho_s} \right)\left(\frac{\vca}{\microt}\right)^2\geq 1 \textrm{,}
\label{ratio}
\end{equation}
where $E_{\microt}\sim \rho_s \microt^2$  is the energy associated with the microturbulent velocity field, $\rho_c$ is the density at the top of the convective region, $\rho_s$ is the surface density and $\microt$ is the microturbulent velocity.
We evaluated the ratio (\ref{ratio}) using the data calculated from our models and a value for the  microturbulent 
velocity $\microt$ of 10$\kms$.
This value has been chosen according to a set of microturbulent velocity determinations in massive stars by {\em Trundle et al. (2007)} and {\em Hunter et al. (2008)}, which has been obtained in the context of the ESO FLAMES Survey of Massive Stars {\em (Evans et al.  2005)}, and has been analyzed in a coherent way. This provides a relatively large data base, even after restricting the data set to slow rotators, i.e. $\vsini < 80\kms$. 
The error in the microturbulent velocity measurements is about 5 $\kms$,  which justifies our choice of $\microt = 10 \kms$ in the evaluation of the ratio (\ref{ratio}).
Fig.~\ref{grwplot}, shows that the process of excitation of microturbulence through sub-surface convection is energetically possible. Moreover, the region where sub-surface convection is efficiently generating gravity waves 
corresponds very well with the location of stars in which a microturbulent velocity field is clearly present.
Using our grid of stellar models and the microturbulent velocity dataset  of   {\em Trundle et al. (2007)} and {\em Hunter et al. (2008)}, these results have been confirmed 
also for stars in the SMC and the Galaxy.
\figureCoAst{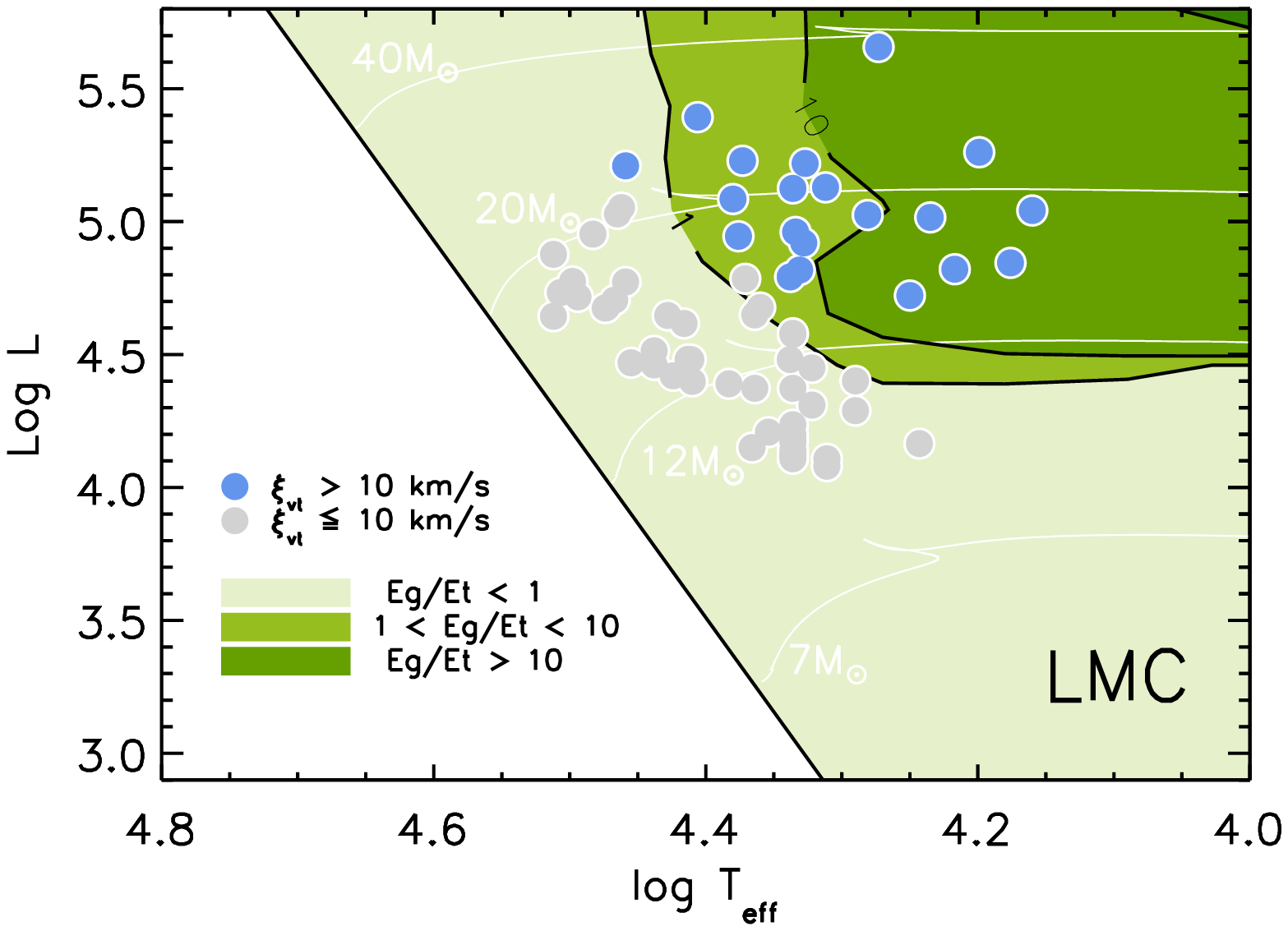}{This HR diagram shows values of the ratio $E_{{\mathrm g}}/E_{\microt}$  for $\microt = 10\kms$ at LMC metallicity. 
  We estimated the ratio as in Eq.~\ref{ratio} using the parameters obtained by stellar evolution calculations. Solid white lines are reference evolutionary tracks.
The full drawn black line is the zero age main sequence.
Over-plotted as filled circles are photospheric microturbulent velocities $\microt$ derived in a consistent way 
for hot massive stars by {\em Trundle et al. (2007)} and {\em Hunter et al. (2008)}. Here, we use only data
for stars with an apparent rotational velocity of $\vsini < 80 \kms $. The uncertainty
in the determination of $\microt$ is typically $\pm 5\kms$. }{grwplot}{t}{clip,angle=0,width=115mm}
 \section*{Clumping at the base of the stellar wind}
 Evidence has been accumulating that the winds of massive stars are 
inhomogeneous on different spatial scales.
Indeed, evidence that O star winds are clumped is given by, among others,
{\em Puls et al. (2006)}. These authors investigate the clumping
behavior of the inner wind (inside about two stellar radii) relative
to the clumping in the outer wind (beyond tens of stellar radii)
of a large sample of supergiant and giant stars.
They find that in stars that have strong winds, the inner 
wind is more strongly clumped than the outer wind, whereas those 
having weak winds have similar clumping properties in the inner and 
outer regions. The analysis only allows for such a relative statement.
Interestingly, this type of radial behaviour is not consistent with
hydrodynamic predictions of the intrinsic, {\em self-excited}
line-driven instability {\em (Runacres \& Owocki 2002,2005)}.
 Such models predict a lower clumping in the
inner wind than in the outer wind.
If we compare the O stars investigated by {\em Puls et al. (2006)} with our models,
the trend is such that stars with relatively strong clumping in the inner winds are in a regime where
the subsurface convective velocity is larger. A correlation between clumping 
at the base of the wind and $\vca$ may point to sub-photospheric
convection as a possible excitation mechanism of clumping at the base
of the wind. 
If mass loss is affected by this clumping, the subsurface convective region in massive stars may impact stellar evolution.

\acknowledgments{Matteo Cantiello acknowledges financial support from HELAS.}

\References{
\rfr Aerts C., et al., this volume
\rfr B\"ohm-Vitense, E. 1958, Zeitschrift fur Astrophysik, 46, 108
\rfr Edmunds, M. G. 1978, A\&A, 64, 103
\rfr Evans, C. J.;, Smartt, S. J., Lee, J.-K., et al. 2005, A\&A, 437, 467
\rfr Goldreich, P. \& Kumar, P. 1990, ApJ, 363, 694
\rfr Hunter I., Lennon, D. J., Dufton, P. L., et al. 2008, A\&A, 479, 541
\rfr Iglesias, C. A. \& Rogers, F. J. 1996, ApJ, 464, 943
\rfr Iglesias, C. A., Rogers, F. J. \& Wilson, B. G. 1992, ApJ, 397, 717
\rfr Lighthill, M. F. 1952, Royal Society of London Proceedings Series A, 211, 564
\rfr Puls, J., Markova, N., Scuderi, S., et al. 2006, A\&A, 454, 625
\rfr Runacres, M. C.  \& Owocki, S. P. 2002, A\&A, 381, 1015
\rfr Runacres, M. C.  \& Owocki, S. P. 2005, A\&A, 429, 323
\rfr Stein, R. F. 1967, Sol. Phys., 2, 385
\rfr Stothers, R. F. \& Chin, C.-W. 1993, ApJ, 408, L85
\rfr Trundle, C., Dufton, P. L., Hunter, I., et al. 2007, A\&A, 471, 625
\rfr Yoon, S.-C., Langer, N., Normann, C. 2006, A\&A, 460, 199
}

\end{document}

%% file: CoAst_liege-proceedingslayo.tex
\renewcommand{\chaptermark}[1]{\markboth{#1}{}}

\newcommand{\vsini}{\ensuremath{v\sin i}}                   
\def\kms {km~s$^{-1}$}

\newcommand{\kopf}{\small\itshape Comm. in Asteroseismology \\ Contribution to the Proceedings of the 38$^{th}$\,LIAC\,/\,HELAS-ESTA\,/\,BAG, 2008
}

\newcommand{\Authors}[1]{\begin{center}\normalsize\bf\sf #1 \end{center}}

\renewcommand{\author}[1]{\begin{center}\normalsize\bf\sf #1 \end{center}}
\newcommand{\Address}[1]{\begin{center}\small\sf #1 \end{center}}

\DeclareGraphicsExtensions{.eps,.jpg}

\newcommand{\Session}[1]{{\vspace{3mm}\small \noindent  \hspace*{3mm} Session: } #1 \normalsize}

	\newcommand{\one}{\small Physics and uncertainties in massive stars on the MS and close to it}

\renewenvironment{abstract}{\section*{Abstract}\normalsize\sf}{}
\newcommand{\References}[1]{\begin{flushleft}{\large References\\}\vspace*{2mm}\small #1 \end{flushleft}}

\newcommand{\chapterCoAst}[2]{\chapter[\sf\normalsize #1\\ \footnotesize \hspace*{5mm}by #2 \sf\normalsize][]{#1\\}\rhead[\fancyplain{}{\sf\footnotesize \center{#1}}]{\fancyplain{}{\sffamily\thepage}}\lhead[\fancyplain{\kopf}{\sffamily\thepage}]{\fancyplain{\kopf}{\sf\footnotesize \center{#2}}}}

\newcommand{\figureCoAst}[5]{\begin{figure}[#4]
\centering
\includegraphics*[#5]{#1}
\caption{#2}
\label{#3}
\end{figure}}

\renewcommand{\chaptername}{}

\newcommand{\acknowledgments}[1]{\vspace*{5mm}\noindent  \textbf{Acknowledgments.} #1}